\documentclass[fleqn,twoside]{article}
\usepackage[headings]{espcrc2}

\readRCS
$Id: espcrc2.tex,v 1.2 2004/02/24 11:22:11 spepping Exp $
\usepackage{graphicx}
\usepackage[figuresright]{rotating}

\newcommand{\AmS}{{\protect\the\textfont2
  A\kern-.1667em\lower.5ex\hbox{M}\kern-.125emS}}

\newcommand{\UtwoS}{\ensuremath{\Upsilon(\mathrm{2S})}}
\newcommand{\UthreeS}{\ensuremath{\Upsilon(\mathrm{3S})}}

\title{ATLAS reach for Quarkonium production and polarization measurements}

\author{Erez Etzion \thanks{erez.etzion@cern.ch} and Jonatan Ginzburg 
\address[TAU]{Tel Aviv University,
Raymond and Beverly Sackler school of Physics and Astronomy. \\
Tel Aviv 69978, Israel \\}
\thanks{This research was supported (in part) by the German-Israeli Foundation for Scientific Reasearch and Development and the Israel Science Foundation}
on behalf of the ATLAS Collaboration}
       
\runtitle{ATLAS reach for Quarkonium measurements}
\runauthor{E. Etzion and J. Ginzburg}

\begin{document}

\begin{abstract}
The ATLAS detector at CERN's LHC is preparing to take data from the first proton-proton collisions expected in the next few
months. We report on the analysis of simulated data samples for production of heavy Quarkonium states J/$\psi$ and $\Upsilon$,
corresponding to an integrated luminosity of 10~$pb^{-1}$ with center of mass energy of 14~TeV expected at the early ATLAS data. 
We review various aspects of prompt Quarkonium production at LHC: the accessible ranges in transverse momentum and pseudorapidity, 
spin alignment of vector states, separation of color octet and color singlet production mechanism and feasibility of observing 
radiative decays $\chi_c$
and $\chi_b$ decays. Strategies of various measurements are outlined and methods of separating promptly produced J/$\psi$
and $\Upsilon$ mesons from various backgrounds are discussed.
\vspace{1pc}
\end{abstract}

\maketitle

\section{Introduction and motivations}
\label{chap:intro}
\begin{figure}[htb]
\vspace{-2 pt}
\includegraphics[width=8.2cm]{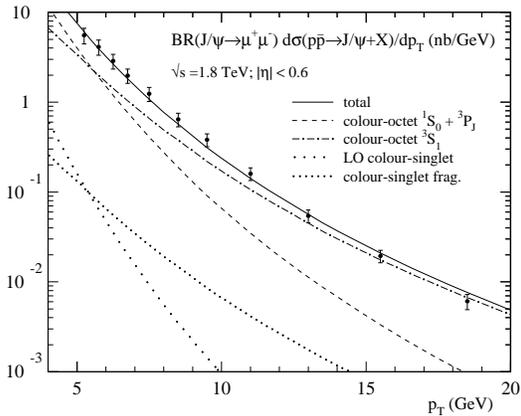}
\caption{Differential cross-section of J/$\psi$ production at CDF with theoretical predictions for color-singlet and color-octet model production\,\cite{kramer-cdf}}
\label{CDF-cross}
\end{figure}

\begin{figure}[htb]
\vspace{-2 pt}
\includegraphics[width=7.8cm]{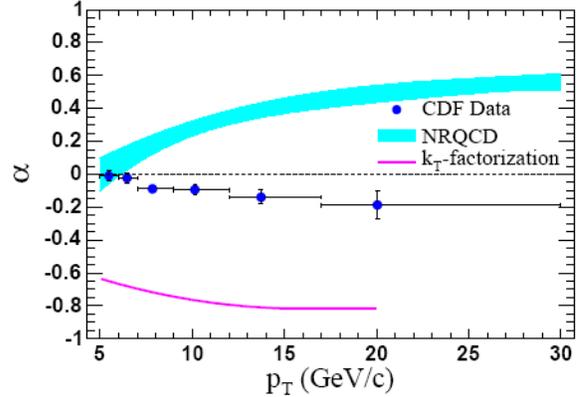}
\caption{Polarizations of J/$\psi$  production as a function of $p_T$ at CDF, (blue circles)
      , curves for limit cases of the $k_T$ factorization model,
      and a band for NRQCD predictions (from \,\cite{Abulencia:2007us}).}
\label{CDF-pol}
\end{figure}

The LHC is expected to produce a large number of quarkonium states such as J/$\psi$ and  $\Upsilon$ already in low luminosity runs during the first few years of running. These events are important for many physics studies. Their sizeable branching fraction into charged lepton pairs allows for easy separation of these events from the expected huge hadronic background.
Being narrow resonances, the quarkonia are perfectly suited for alignment and calibration of the ATLAS trigger and tracking systems.
On top of that, understanding the details of the prompt charmonia production is a challenging task and a good testbed for various QCD calculations, spanning both perturbative and non-perturbative regimes through the creation of heavy quarks in the hard process and their subsequent evolution
into physical bound states. This note concentrates on the capabilities of the ATLAS detector to study the prompt charmonia production and polarization measurement at LHC. 
One approach which can be used to describe the evolution of heavy quark antiquark pair into quarkonium bound state is the Color Evaporation Model (CEM)\,\cite{Amundson-all}. In CEM there is no correlation of color and angular momentum quantum numbers between the initial  $Q\bar{Q}$ state and the final quarkonium state. It is assumed that soft gluon emission during the binding process alters the quantum numbers to arrive at the appropriate final state.
Before the Tevatron experiments started to collect data it was assumed that the Color Singlet Model (CSM)\,\cite{orig-ccbar} is the right explanation for quarkonium production mechanism. This model assumes that each quarkonium state can only be produced by a $Q\bar{Q}$ pair in the same color and angular momentum state as that quarkonium. Therefore, a J/$\psi$ meson, for example, can only be derived from a $c\bar{c}$ pair created in a $1^3S_1$ color-singlet state. The attractive feature of the CSM is that it makes definite quantitative predictions for production cross sections, because the binding probability is related to the quarkonium wavefunction from the potential model. However, this approach predicted the J/$\psi$ production rate an order of magnitude lower than that measured by CDF.
The Nonrelativistic QCD (NRQCD) Colour Octet Model (COM)\,\cite{Bodwin-Braaten} was proposed to explain this discrepancy.
In this effective field theory $Q\bar{Q}$  pairs produced with one set of quantum numbers can evolve into a quarkonium state with different quantum numbers, by emitting low energy gluons. In contrast to the CSM, in which such transitions have zero probability. The good description of the Tevatron data by the COM model shown in Figure~\ref{CDF-cross} is at least in part due to the tuning of some of it's parameters which were determined from the same data. However, the difference between the CDF measurement of J/$\psi$ polarization dependence on transverse momentum ($p_T$) and the theory predictions (Figure~\ref{CDF-pol}) is motivating us to repeat the measurement in the LHC higher energy regime utilizing the advantage ATLAS will have in the number of events and the higher $p_T$ reach.
\section{ATLAS di-muon trigger}
\label{sec:trigger}
\begin{figure}[htb]
\vspace{-2 pt}
\includegraphics[width=8.2cm]{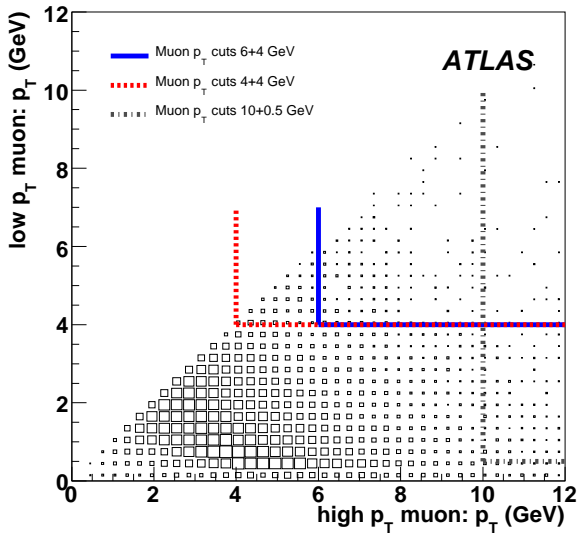}
\caption{Density of J/$\psi$ production cross-section as a function of the hardest and 
      softer muon $p_T$ of muons coming from J/$\psi$. No cut was placed on the generated sample, 
      but the overlaid lines represent the thresholds of observed events with trigger cuts applied.}
\label{fig:Jpsilowpttrigger}
\end{figure}

The LHC will collide two 7~TeV proton beams at a rate of 40~MHz, which together with pile-up will result in an interaction rate of up to around 1~GHz. The ATLAS trigger acts to reduce this rate to around 100~Hz written to disks, whilst keeping only events `of interest'.
The trigger first level (LVL1) decision is based on  coarse granularity of two sub-detector systems: the muon trigger chambers and the calorimeters. 
The second level trigger (LVL2) receives from LVL1 data restricted to limited Regions of Interest (RoI). For a LVL1 muon, the LVL2 will use the information from the muon and inner detector tracking chambers to improve the muon momentum estimate, which allows a tighter selection based on this quantity.
There are two specific types of trigger dedicated to quarkonium: one which requires two LVL1 RoIs corresponding to two muon candidates 
with $p_T$ values above thresholds of 4 and 6~GeV respectively, and the other requires a single LVL1 RoI above a threshold of 4~GeV and searches for the second muon of opposite charge in a wide RoI at LVL2. We consider a complementary trigger on J/$\psi$ events triggering on a single higher $p_T$ (10~GeV) muon and searching for a matched track at the off-line analysis level. Before incorporating trigger and reconstruction efficiencies, the predicted cross-sections
for $pp\rightarrow J/\psi\rightarrow\mu^+\mu^-+X$ were calculated for a number of $p_T$ thresholds on the di-muon trigger (see Table~\ref{tab:oniaxsecs}
for details). Figures~\ref{fig:Jpsilowpttrigger} illustrates the distribution of
cross-sections across the values of the $p_T$ of the harder and softer muon from the quarkonium decay
without any muon cuts applied and zero polarization. 
The lines overlaid on the plots represent the the muon trigger thresholds $p_T>6$~GeV and $p_T>4$~GeV for the harder and the software muon respectively (denoted further by "6+4~GeV" or  $\mu6\mu4$),
similarly "4+4~GeV" ($\mu4\mu4$)  refers to the trigger threshold $p_T>4$~GeV applied on both muons and "10+0.5~GeV" ($\mu10$) refers to trigger $p_T$ threshold of 10~GeV applied on one of the muons only. In all cases a pseudorapidity, $\eta$, of a muon lies within an interval $|\eta|<2.5$.

Even if the bulk of the J/$\psi$ are produced under the trigger thresholds, nevertheless, due to the high cross-section the number of accepted J/$\psi$ will be substantially larger than at the Tevatron.
In the $\Upsilon$ case, due to it's relatively larger mass the bulk of the production is in the region near the muon threshold of 5 and 4~GeV.
This means that by lowering the threshold for the higher momentum muon we can significantly increase the number of recorded $\Upsilon$ events.

Estimated quarkonia cross-sections for the three scenarios are presented in Table~\ref{tab:oniaxsecs}. The production will be dominated by J/$\psi$ and $\Upsilon(1S)$. 
while $\psi^\prime$ and $\Upsilon(2S)$, $\Upsilon(3S)$ are expected to give smaller contributions.

\begin{table}[htb]
\caption{Predicted cross-sections for various prompt vector quarkonium state production
  and decay into muons, with di-muon trigger thresholds $\mu4\mu4$ and $\mu6\mu4$
  and the single muon trigger threshold $\mu10$ (before trigger and reconstruction efficiencies).
  The last column shows the overlap between the di-muon and single muon samples.}
  \label{tab:oniaxsecs}
\begin{tabular}{|c||c|c|c|| c |}
\hline
{Quarkonium}  & \multicolumn{4}{c|}{Cross-section, nb} \\
                              & $\mu4\mu4$ & $\mu6\mu4$ & $\mu10$ & $\mu6\mu4$\\
                              &           &            &          & and $\mu10$ \\
\hline
J/$\psi$           & 28  & 23  & 23   &  5   \\
$\psi^\prime$   & 1.0 & 0.8 & 0.8  &  0.2    \\
$\Upsilon(1S)$          & 48  & 5.2 & 2.8  &  0.8 \\
\UtwoS          & 16  & 1.7 & 0.9  &  0.3    \\
\UthreeS        & 9.0 & 1.0 & 0.6  &  0.2   \\
\hline
\end{tabular}
\end{table}

\begin{figure}[htb]
\vspace{-2 pt}
\includegraphics[width=8.0cm]{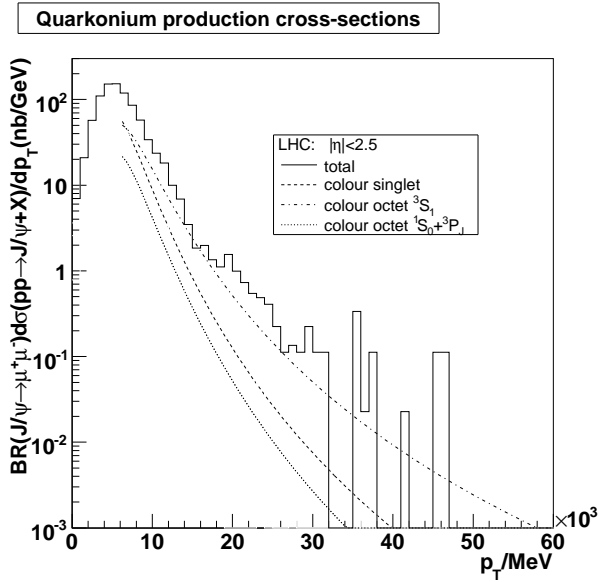}
\caption{Monte Carlo differential cross-section for J/$\psi$ production with J/$\psi$ $p_T$ with 
      the differential cross-section contributions from direct color singlet (dashed line), $c\bar{c}$ states contribute to J/$\psi$ through $\chi$ decays 
      (dotted line) and octet production (dashed dotted line) overlaid.}
\label{fig:charmMCmuoncuts}
\end{figure}

The charmonium production is composed of three main processes: direct singlet production, octet production and singlet production of $\chi$
states. Each of these processes is characterized by different differential cross-sections, $\frac{d\sigma}{dp_T}$.
Figure~\ref{fig:charmMCmuoncuts} illustrates the contributions of these three 
classes to the overall production rate for J/$\psi$ once muon $p_T$ trigger cuts of 6 and 4~GeV 
are applied to the muons coming from the J/$\psi$.

\section{Charmonium reconstruction}
\label{sec:charmonium}

\begin{figure}[htbp]
  \includegraphics[width=8cm]{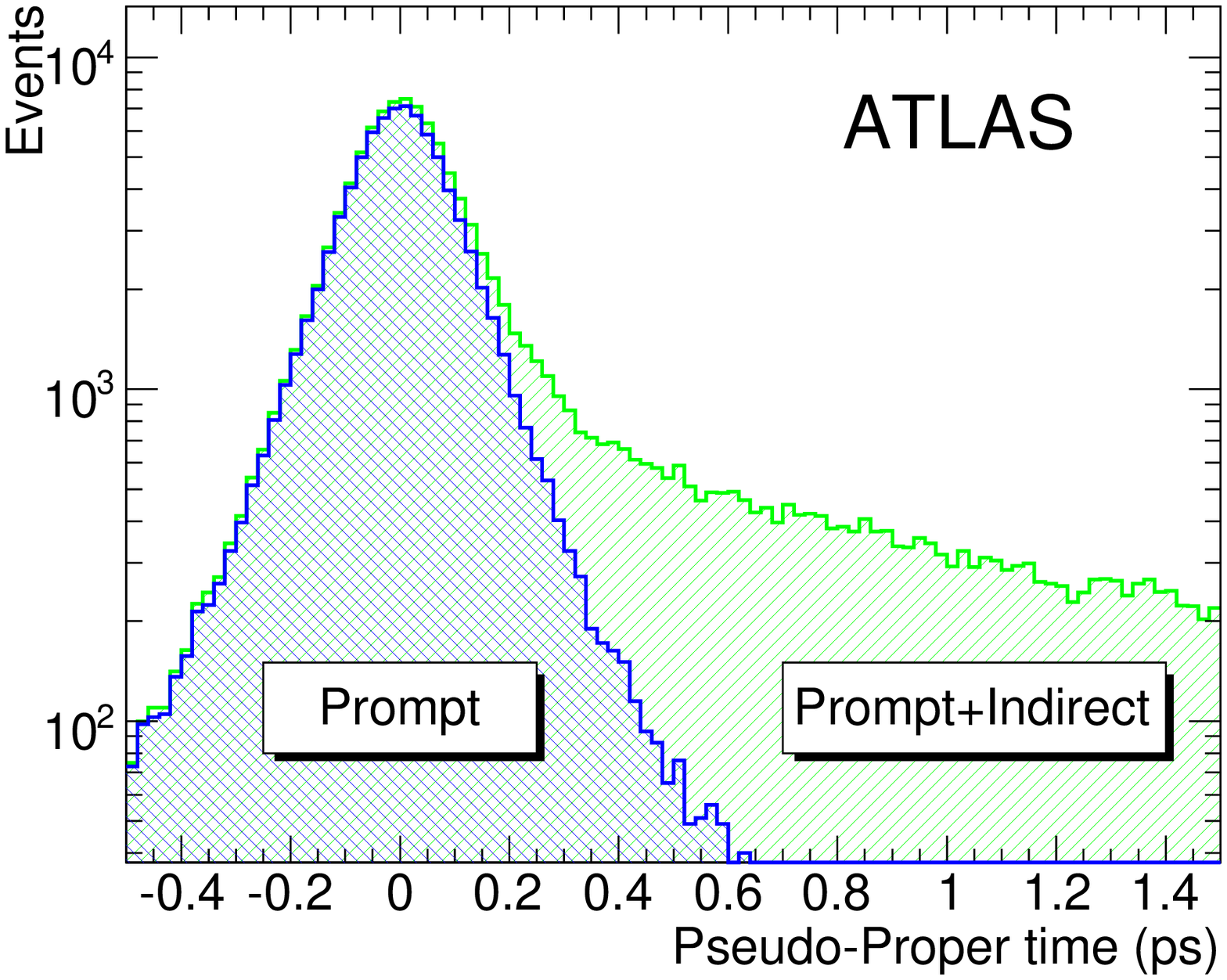}
  \includegraphics[width=8cm]{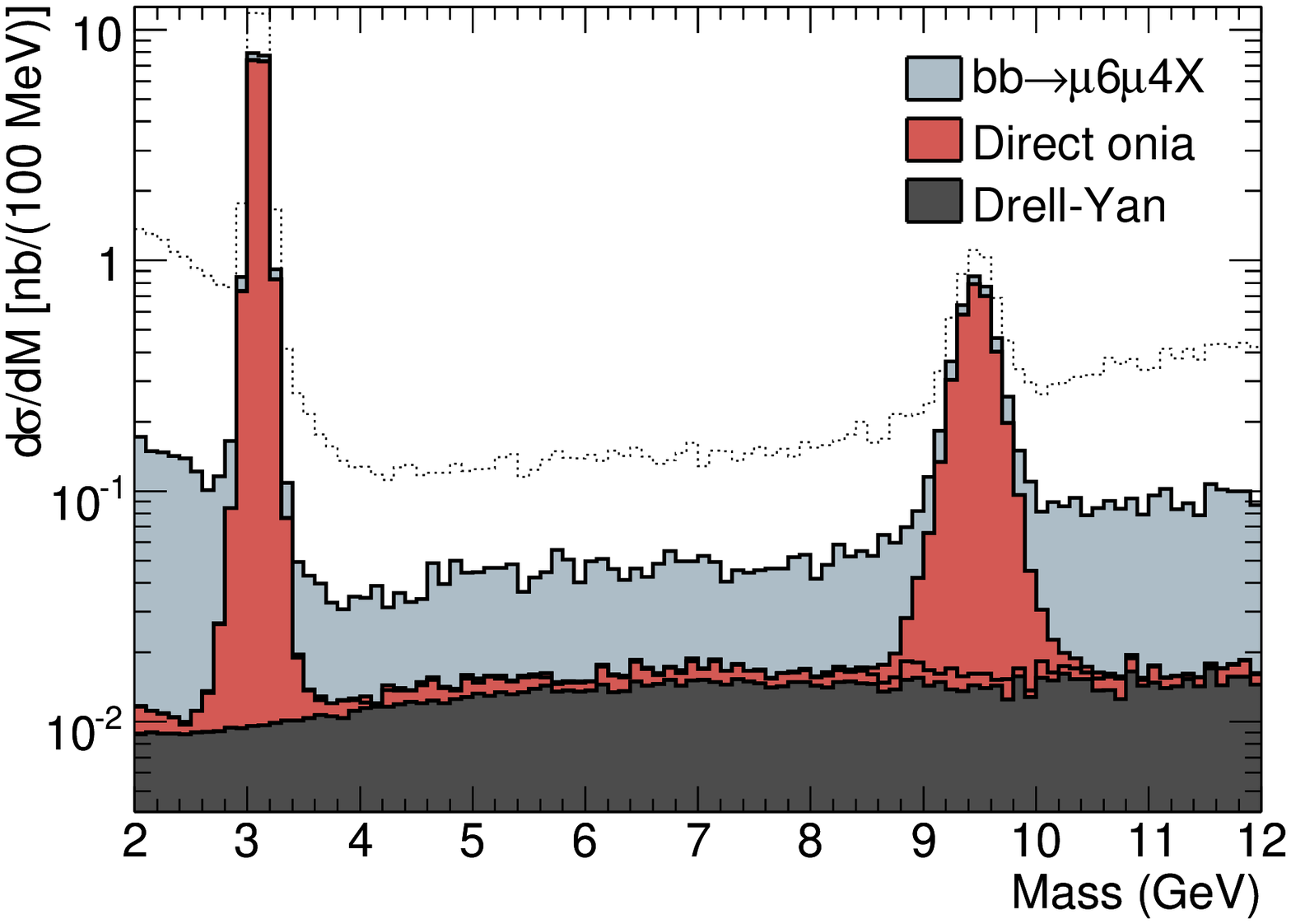}
\hfill\\
  \caption{top - Pseudo-proper decay time, $\tau$,  distribution for reconstructed prompt J/$\psi$
      (cross-hatched, centred at zero) and the sum of prompt and indirect J/$\psi$ from B-decays
      (hatched, exponential distribution).\newline
      bottom -  Sources of low invariant mass di-muons, reconstructed
      with a $\mu6\mu4$ $p_T$ trigger, with the requirement
      that both muons are identified as coming from a primary vertex
      and with a pseudo-proper decay time smaller than 0.2~ps. The white area represent the background that was rejected with the pseudo-proper decay time treatment.}.
    \label{fig:jpsipropertime}
\end{figure}

In any event which passes the di-muon trigger, all the reconstructed muon candidates are combined into oppositely charged pairs, 
and each of these pairs is analyzed in turn. 
If the invariant mass of the two muons is above 1~GeV, we attempt to refit the tracks to a common vertex.
The fraction of surviving J/$\psi$ candidates before the refit is 96\%.
If a good vertex fit is achieved, the pair is accepted for further analysis. 
If the invariant mass of the refitted tracks is within 300~MeV for J/$\psi$ or 1~GeV for $\Upsilon$ (six times the expected average mass resolution) of the 
expected mass, the pair is considered as a charmonium candidate.
The reconstructed efficiency varies with $p_T$ and $\eta$. When the J/$\psi$ $p_T>10$~GeV we get a sharp rise in its acceptance. The rise in the $\Upsilon$ case is less sharp and it reaches a high acceptance level around 30~GeV. Both channels reach a similar plateau at around acceptance of 85\%. Due to the trigger requiring both muons to pass the trigger $p_T$ threshold, the angular separation between the two muons is not large for most of the accepted events. Describing the opening angle by $\delta R=\sqrt{\delta\phi^2+\delta\eta^2}$, where $\delta\phi$ and $\delta\eta$ are the differences of the azimuthal angle and the  of the two muons from J/$\psi$.  Typical values for $\delta R$ are $\approx 0.47$ which means that requiring both muons to be above the $p_T$ threshold forces them to fly very close one to each other.
For that reason the J/$\psi$ angular acceptance follows closely the individual muon distribution and its dependence on material and detector effects. In the $\Upsilon$ case, due to its higher mass, it tends to be produced with higher $p_T$. For that reason the separation angle $\delta R$ is much wider, the muons do not go to the same area in the detector and therefore the efficiency dependence on pseudorapidity is much smoother than in the J/$\psi$ case.
The main sources of low invariant mass di-muons expected to dominate the background for prompt charmonium are:
\begin{itemize}
\item Decays in flight of $\pi^\pm$ and $K^\pm$ - muons from this channel have a steeply falling momentum spectrum. Their contribution in the mass region of interest
is expected to be very small where the requirement for random coincidences
in the charmonium invariant mass range reduces it to below the 1\% level.
\item Di-muon production via the Drell-Yan process - only small fraction of these events survive the di-muon trigger requirements, which makes this background essentially negligible. 
\item Continuum of muon pairs from beauty (and charm) decays - contributions from charm decays have not been simulated, but despite having
an estimated total rate twice as high as from beauty for a $\mu6\mu4$ trigger,
as the $p_T$ distributions of muons from charm quarks falls more steeply,
charm sources are expected to contribute at a lower level than beauty.
\item Indirect J/$\psi$ production - this is the main source of expected background, its reduction is discussed below.
\end{itemize}
All the background sources apart from the Drell-Yan pairs contain muons which do not originate from the interaction point, this is used to suppress their contamination by rejection of the events containing a secondary vertex, if identified.
The radial displacement of the two-track vertex from the beamline is used to distinguish
between prompt J/$\psi$ and B-hadron decays having an exponentially decaying
pseudo-proper time distribution due to the lifetime of the 
B-hadrons that may decay into quarkonia. The pseudo-proper decay time, $\tau$, is defined as
\begin{center}
\begin{equation}
\tau = \frac{L_{xy}\cdot M_{J/\psi}}{p_T(J/\psi)},
\end{equation}
\end{center}
where $M_{J/\psi}$ and $p_T(J/\psi)$ represent the J/$\psi$  invariant mass and
transverse momentum, and $L_{xy}$
is the transverse decay length of the meson. The resolution in the pseudo-proper decay time is expected to vary from 0.110~ps for the low $p_T$ charmonia down to 0.07~ps for the higher $p_T$ lighter angular dependence is very vague. 
As demonstrated in  Figure~\ref{fig:jpsipropertime} a cut on this quantity can efficiently distinguish between prompt and indirect J/$\psi$ events.

Using the $\mu10$ trigger each reconstructed single muon candidate
is combined with oppositely-charged tracks reconstructed in the same event within a cone of $ \delta R = 3.0$. Any track, including those that were not identified as muons, are examined.
As in the di-muon analysis, we require that both the identified muon and the
track are flagged as having come from the primary vertex. In addition,
 we impose a cut on the transverse impact parameter $d_0$, 
$|d_0|<0.04$~mm on the identified muon and $|d_0|<0.10$~mm on the second track, in order to further suppress 
the number of background pairs from $B$-decays.
We obtain a J/$\psi$  invariant mass resolution close to that in the di-muon sample.
It's worth noting that the signal-to-background ratio
around the J/$\psi$ peak improves slightly with increasing transverse
momentum of J/$\psi$.
At higher $p_T$ the $\cos\theta^\ast$ acceptance also becomes 
broader, which should help independent polarization measurements.
\section{Polarization studies}
\label{sec:polarization}
The polarizations of  quarkonia can be measured using the angular distribution 
of the daughter particles produced in the decay. 
\begin{center}
\begin{equation}
\label{eqn:spinalign}
  \frac{d\Gamma}{d\cos\theta^\star}\propto 1+\alpha \cos^2 \theta^\star,
\end{equation}
\end{center}
where  $\cos\theta^\star$ is the angle between the direction
of the positive (by convention) muon from 
quarkonium decay in the quarkonium rest frame and the direction of quarkonium itself in the laboratory
frame.
The polarization parameter $\alpha$, defined as 
$\alpha=(\sigma_T-2\sigma_L)/(\sigma_T+2\sigma_L)$, is equal to $+1$ for 
transversely polarized quarkonia production, (helicity $\pm1$). $\sigma_T$ and $\sigma_L$ are the transverse and longitudinal cross sections. For longitudinal (helicity 0) polarization $\alpha$ is equal 
to $-1$. Unpolarized production consists of equal fractions of helicity states 
$+1$, $0$ and $-1$, and corresponds to $\alpha=0$. 

The previous Tevatron measurements were limited to below around 20~GeV
where the polarization is best predicted and the theory most understood.
At ATLAS we aim to measure the polarization of directly produced 
prompt quarkonium in the $p_T$ region up to
$\sim 50$~GeV and beyond with extended coverage in $\cos\theta^\star$. This
will allow for improved fidelity of efficiency,
better discrimination of longitudinal and transverse polarizations and therefore reduced systematics uncertainties.

Two methods for polarization measurements have been examined. In the first one we used a templates chi-squared fit to the $cos^2\theta^\star$ distribution in  six $p_T$ bins of 
 Monte Carlo (MC)-data sample. A linear combination of three MC generated templates histograms are fitted to the angular distribution of the data. The first is for a longitudinally polarized sample the second for a transversely polarized sample and the third is for background events sample. 

\begin{figure}[htb]
  \begin{center}
    \includegraphics[width=8.0cm]{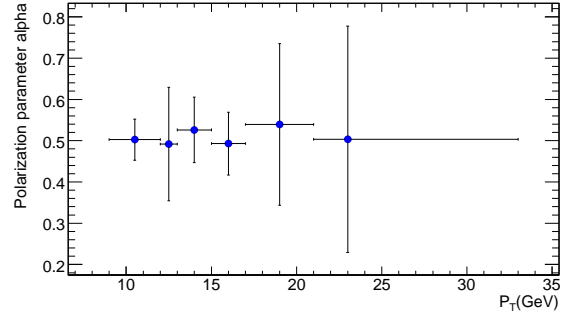}
    \caption{
     The measured polarization parameter $\alpha$ and its error in six $p_T$ bin. The initial polarization of $\alpha=0.5$ was set to the MC data sample.
	 Statistics corresponding to integrated luminosity of about 5~pb$^{-1}$.
      \label{fig:PolarizationAlpha}
    }
  \end{center}
\end{figure}

Feed-down from $\chi$ states and B-decays dilute the prompt
sample and lead to an effective depolarization which is difficult to measure.
The background events sample was generated with zero polarization ($\alpha=0$) using the same detector acceptance effects.
The MC samples generated are  large enough, and as a result the statistical fluctuations in the templates are negligible.
The measured values of $\alpha$ in six bins of $p_T$ are presented in Figure~\ref{fig:PolarizationAlpha}.
As one can see from this plot the average polarization is well concentrated around $\alpha=0.5$, which is the initial value of polarization 
that was set in MC-data sample.

\begin{figure}[htbp]
     \includegraphics[width=8.0cm]{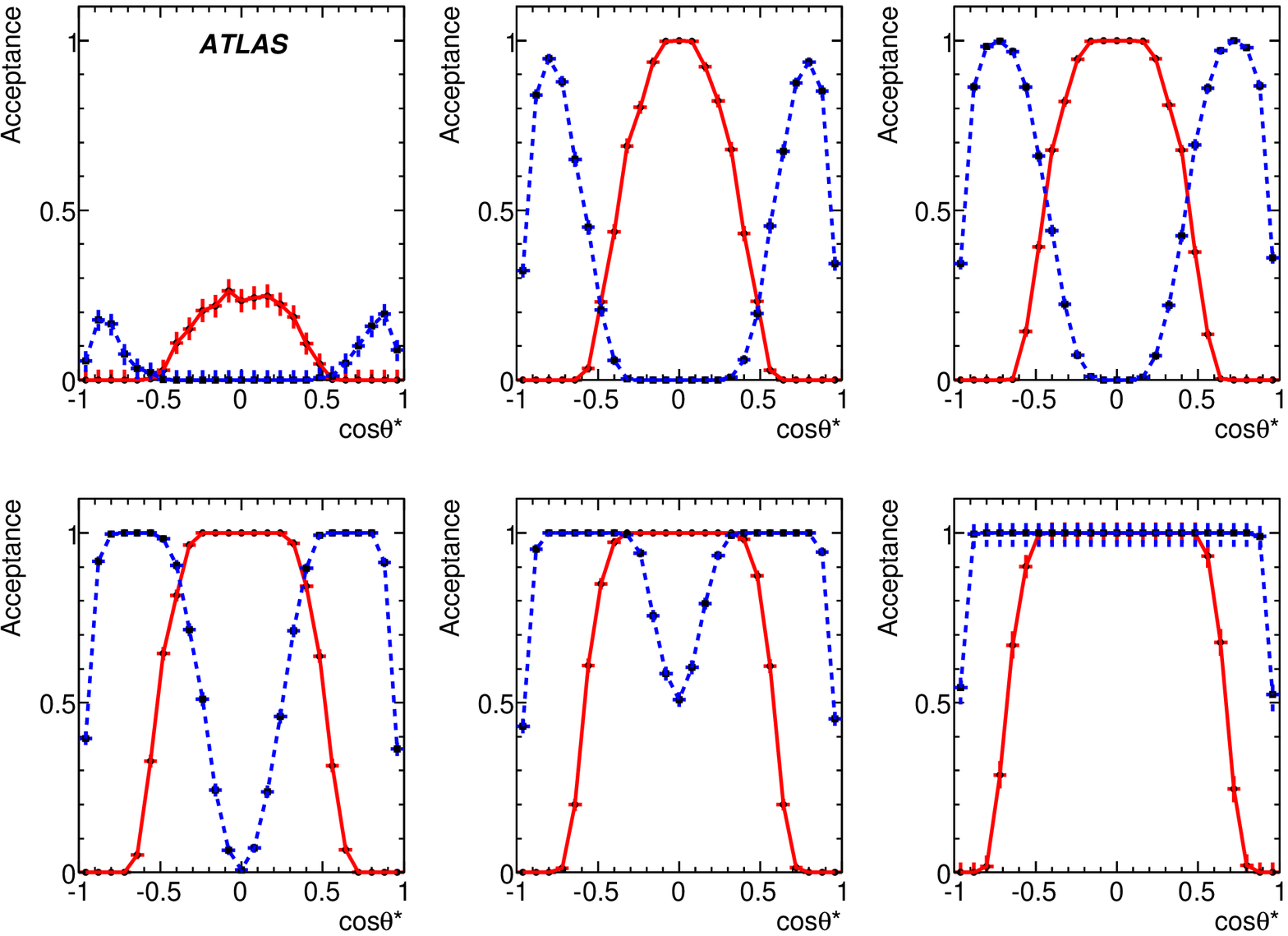}
     \includegraphics[width=8.0cm]{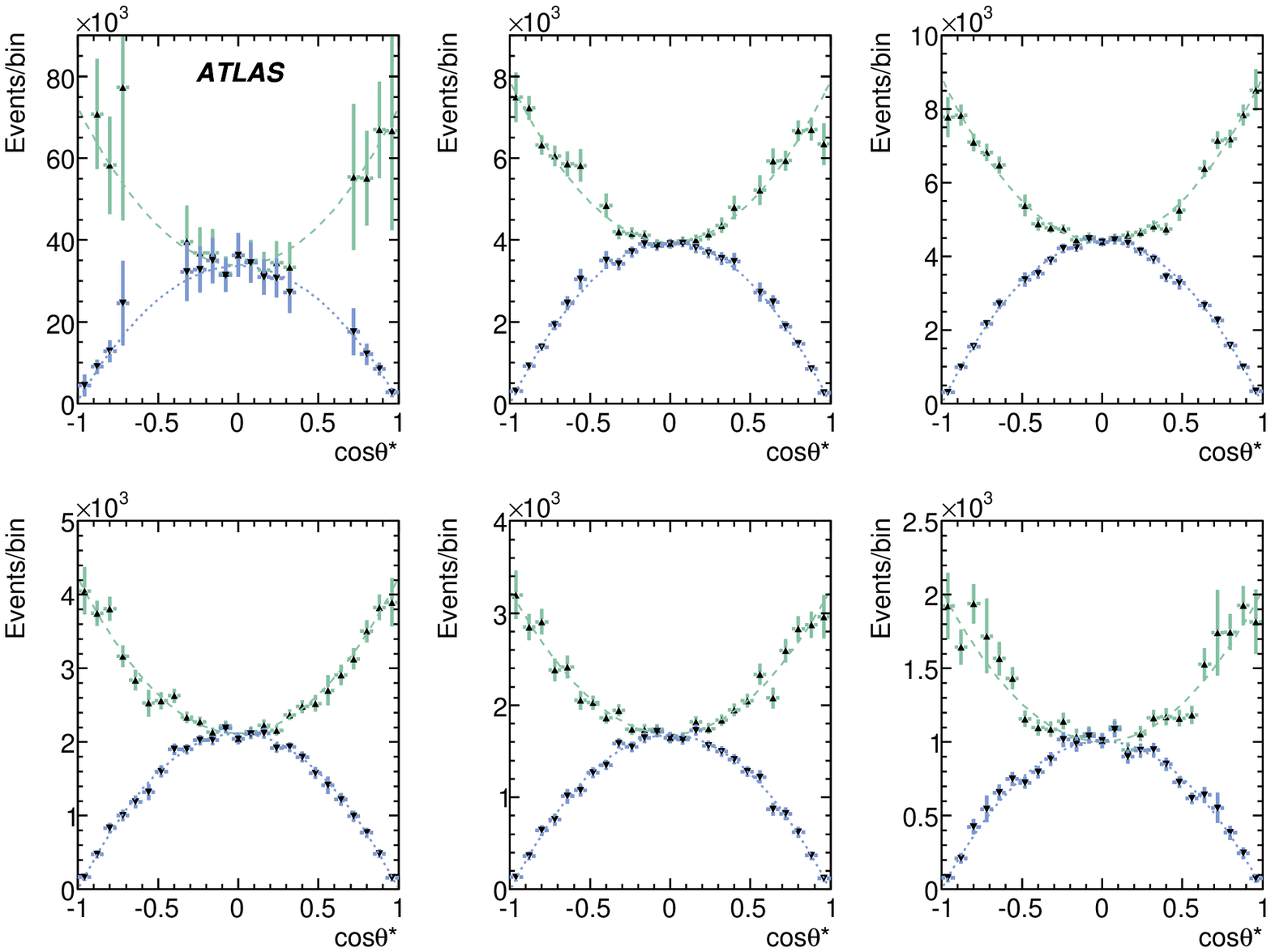}
\hfill\\ 
   \caption{Six figures at the top show kinematic acceptances of the $\mu6\mu4$ (solid red line)
     and $\mu10$ (dashed blue lines) generator level cuts, 
     calculated with respect to the sample with no generator level 
     cuts on muon $p_T$, in slices
     of J/$\psi$ transverse momentum: left to right, top to bottom 9-12
     12-13, 13-15, 15-17, 17-21, above 21~GeV.\newline
     Six bottom figures show combined and corrected (equation ~\ref{eq:measure}) distributions in polarization angle, for longitudinally (dotted lines) and transversely (dashed lines) polarized J/$\psi$ sample in the same $p_T$ slices with statistics corresponding to integrated luminosity of  10~pb  $^{-1}$.}
      \label{fig:jpsimu6mu4-polarisation-ptslices}
\end{figure}

Evidently polarization measurement can significantly suffer from low $|\cos\theta^\star|$ acceptance, and hence from difficulties 
in separating detector efficiency corrections from polarization state effects.
The di-muon trigger requirement for both muons to be above a certain $p_T$ threshold ($\mu6\mu4$) reduces to minimum the acceptance at 
large values of $|\cos\theta^\star|$, where the  difference between various polarization states is the more pronounced (see for example the two polarization states on the lower plots of Figure~\ref{fig:jpsimu6mu4-polarisation-ptslices}).
The acceptance of the single muon trigger sample ($\mu10$) is very different. Here the efficiency is higher at large values of $|\cos\theta^\star|$ and drops in around zero. Figure~\ref{fig:jpsimu6mu4-polarisation-ptslices} (on the top) demonstrates how the two samples complement each other mainly at the low $p_T$ regions, while at high $p_T$ the two triggers increasingly overlap, thus allowing for a cross-check of acceptance and efficiency corrections.
In order to achieve this, the $p_T$ distributions of both samples, $\mu6\mu4$ and $\mu10$ events,
were appropriately  combined.
The combined distribution $dN^{\mathrm{raw}}/d\cos\theta^\ast$, 
was corrected according to equation~\ref{eq:measure}, 
\begin{eqnarray}
\frac{dN^{\mathrm{cor}}}{d\cos\theta^\ast} =
\frac{1}{{\cal{A}}(p_T,\cos\theta^\ast)
\cdot \varepsilon_1
\cdot \varepsilon_2}
\cdot\;
\frac{dN^{\mathrm{raw}}}{d\cos\theta^\ast}.
\label{eq:measure}
\end{eqnarray}
Here $\varepsilon_1$ stands for the trigger and reconstruction
efficiency, while $\varepsilon_2$ denotes the efficiency of
background suppression cuts for each sample, and ${\cal{A}}(p_T,\cos\theta^\ast)$ is the kinematic acceptance of the triggers selection.
These $\cos\theta^\ast$ distributions are fitted using 
equation~\ref{eqn:spinalign}, with $\alpha$ and normalization as free parameters
for each $p_T$ slice. 
The results obtained when fitting unpolarized sample ($\alpha=0$) with statistics corresponding to 10~pb$^{-1}$ are:
$\alpha=0.156\pm0.166, -0.006\pm0.032, 0.004\pm0.029, -0.003\pm0.037, -0.039\pm0.038$ and $0.019\pm0.057$ corresponding to the $p_T$ bins as in Figure~\ref{fig:jpsimu6mu4-polarisation-ptslices}. The precision of the cross section derived from the normalization factor are $\pm 4.35$ in the first slice and decreasing from $\pm0.09$ to $\pm0.04$ with increasing $p_T$ slices. Repeating the same study with $\Upsilon$ we get back numbers that are consistent with ($\alpha=0$) but with larger errors running from $\pm0.17$ to $\pm0.22$.
To further check the ability to measure the polarization we reweighted the raw MC distribution to emulate transversely polarized ($\alpha=+1$) and longitudinally polarized ($\alpha=-1$) J/$\psi$ samples. The same analysis was repeated and the results can be seen on the bottom of Figure~\ref{fig:jpsimu6mu4-polarisation-ptslices}

\section{Analysis of $\chi$ production}
\label{sec:chireco}

A sizeable fraction of prompt J/$\psi$ and $\Upsilon$ 
are expected to originate from radiative decays of heavier states,
$\chi_c$ and $\chi_b$.
These states have even $C$ parity and therefore have a strong coupling to
the color-singlet two gluon state.
About 30 to 40~\% of J/$\psi$ in our signal will come from decays of $\chi_c \rightarrow J/\psi +\gamma$.
Unfortunately, the energies of these photons tend to be quite small.
The ability of ATLAS to detect these photons and resolve various $\chi$
states is rather limited.
Trying to evaluate that for each reconstructed charmonium candidate 
we calculated the invariant mass of the $\mu\mu\gamma$ system with all the photons found in that event.
The  $\mu\mu\gamma$ system is considered to be a $\chi$ candidate, if:
the difference between the invariant masses of the $\mu\mu\gamma$ and
$\mu\mu$ systems lies between 200 and 800~MeV, and
the cosine of the opening angle between the
J/$\psi$ and $\gamma$ is larger than 0.98.

We fitted three guassians to the difference in invariant masses of the $\mu\mu\gamma$ and
$\mu\mu$ measured in those $\chi$ candidates. The three mean positions were fixed according to the 318, 412 and 460~MeV corresponding to the $\chi_0$, $\chi_1$ and $\chi_2$ expected values. 
The fit parameters were the heights of the
three Gaussian peaks $h_{\chi0}, h_{\chi1}, h_{\chi2}$, and the 
three parameters describing the smooth polynomial background of J/$\psi$ production from B-hadron decays which survive the pseudo-proper decay time requirement.
The MC input amplitudes of the peaks (15, 123 and 87, respectively)
were reproduced reasonably well:
$h_{\chi0} = 15 \pm 3(stat) \pm 10(sys), h_{\chi1} = 101 \pm 4(stat) \pm 12(sys), h_{\chi2} =103 \pm 4(stat)  \pm9(sys)$
with strong negative correlation between the last two. The overall $\chi_c$ reconstruction efficiency is estimated to be about 5\%.

\section{Physics reach with early ATLAS data}
\label{sec:summary}

During the initial run of the LHC with the luminosity of $10^{31}$~cm$^{-2}$~s$^{-1}$, one day of running corresponds to integrated luminosity of 1~pb$^{-1}$. This can be translated to about di-muon triggered 15,000 J/$\psi \rightarrow\mu\mu$ and 2,500 of $\Upsilon \rightarrow\mu\mu$. To that we can add the sample of 16,000 and 2,000  J/$\psi$ and $\Upsilon$ triggered with one muon above 10~GeV. In 10 days we expect to collect a sample which is roughly equal to the MC statistics used in this note which may be enough for the first cross section and polarization $p_T$ dependence measurements. The precision of the J/$\psi$ polarization can reach 0.02-0.06 depending on the level of polarization itself, as well as on our understanding the trigger and detector efficiencies, resolution and background estimations.\newline
By the end of one year, with the expected integrated luminosity
of 100~pb$^{-1}$, the transverse momentum spectra are expected
to reach about 100~GeV and possibly beyond.
With several million J/$\psi\rightarrow\mu\mu$ decays, and better understanding
of the detector, $\chi_c\rightarrow J\/\psi\gamma$ should become observable,
while other measurements mentioned above will become increasingly precise.
During the future high luminosity runs, we will have to increase the trigger threshold, and rescale the lower trigger threshold.
Nevertheless the higher luminosity will further expand the range of
reachable transverse momenta, and allow further tests of the production mechanisms and the polarization measurements,
as well as make $\chi_c$ reconstruction easier.





\begin{thebibliography}{00}




\bibitem{Amundson-all} 
J.F. Amundson et al., "Quantitative Tests of Color Evaporation: Charmonium
Production", Phys. Lett. B390, 323 (1997).

\bibitem{orig-ccbar}
  See e.g.
  V.~G.~Kartvelishvili, A.~K.~Likhoded, S.~R.~Slabospitsky,
  Sov.\ J.\ Nucl.\ Phys.\  {\bf 28} (1978) 280;
  M.~Gluck, J.~F.~Owens and E.~Reya,
  ``Gluon Contribution To Hadronic J/$\psi$ Production,''
  Phys.\ Rev.\ D {\bf 17} (1978) 2324;
  V.~G.~Kartvelishvili, A.~K.~Likhoded,
  Sov.\ J.\ Nucl.\ Phys.\  {\bf 39} (1984) 298.

\bibitem{Bodwin-Braaten} 
G.T. Bodwin, E. Braaten , G.P. Lepage, "Rigorous QCD Analysis of Inclusive
Annihilation and Production of Heavy Quarkonium", Phys. Rev. D 51, 1125
(1995), Erratum ibid. D55, 5853 (1997).

\bibitem{kramer-cdf}
  M.~Kramer,
  ``Quarkonium production at high-energy colliders,''
  Prog.\ Part.\ Nucl.\ Phys.\  {\bf 47} (2001) 141
  [arXiv:hep-ph/0106120].

\bibitem{Abulencia:2007us}
  A.~Abulencia {\it et al.}  [CDF Collaboration],
  ``Polarization of $J/\psi$ and $\psi_{2S}$ mesons produced in $p \bar{p}$
  collisions at $\sqrt{s}$ = 1.96-TeV,''
  Phys.\ Rev.\ Lett.\  {\bf 99} (2007) 132001
  [arXiv:0704.0638 [hep-ex]].
 
\end{thebibliography}
\end{document}